\documentclass{jfm}
\usepackage{epsf}

\ifCUPmtlplainloaded \else
  \checkfont{eurm10}
  \iffontfound
    \IfFileExists{upmath.sty}
      {\typeout{Found AMS Euler Roman fonts on the system,
                   using the 'upmath' package.}%
       \usepackage{upmath}}
      {\typeout{Found AMS Euler Roman fonts on the system, but you
                   dont seem to have the}%
       \typeout{'upmath' package installed. JFM.cls can take advantage
                 of these fonts, if you use 'upmath' package.}%
       \providecommand\upi{\pi}%
      }
  \else
    \providecommand\upi{\pi}%
  \fi
\fi

\ifCUPmtlplainloaded \else
  \checkfont{msam10}
  \iffontfound
    \IfFileExists{amssymb.sty}
      {\typeout{Found AMS Symbol fonts on the system, using the
                'amssymb' package.}%
       \usepackage{amssymb}%
         \let\leq=\leqslant
         \let\geq=\geqslant
      }{}
  \fi
\fi

\ifCUPmtlplainloaded \else
  \IfFileExists{amsbsy.sty}
    {\typeout{Found the 'amsbsy' package on the system, using it.}%
     \usepackage{amsbsy}}
    {\providecommand\boldsymbol[1]{\mbox{\boldmath $##1$}}}
\fi

\newcommand\Rey{\mbox{\textit{Re}}}  
\newcommand\Ei{\mbox{Ei}}            
\newcommand\Ci{\mbox{Ci}}            
\newcommand\Si{\mbox{Si}}            
\newcommand\ue{\mathrm{e}}           
\newcommand\uexp{\mathrm{exp}}       
\newcommand\ui{\mathrm{i}}           
\newcommand\ucc{\mathrm{c.c.}}      
\newcommand\uhot{\mathrm{h.o.t.}}   
\newcommand\ud{\mathrm{d}}          
\newcommand\usign{\mathrm{sign}}    
\newcommand\usin{\mathrm{sin}}      
\newcommand\ucos{\mathrm{cos}}      
\newcommand\bu{\boldsymbol{u}}
\newcommand\p{\ensuremath{\partial}}

\newcommand\eps{\epsilon}
\newcommand\Om{\Omega}
\newcommand\om{\omega}
\newsavebox{\astrutbox}
\sbox{\astrutbox}{\rule[-5pt]{0pt}{20pt}}
\newcommand{\astrut}{\usebox{\astrutbox}}

\title[Flow induced by a randomly ...]
{Flow induced by a randomly vibrating boundary}
\author[D. Volfson and J. Vi\~nals]{DMITRI VOLFSON$^{1}$ AND JORGE 
VI\~NALS$^{1,2}$}

\affiliation{$^1$Supercomputer Computations Research Institute,
             Florida State University, Tallahassee, Florida 32306-4130, USA\\
             $^{2}$ Department of Chemical Engineering,
             FAMU-FSU College of Engineering, Tallahassee, Florida
                31310-6046, USA}
\pubyear{1999}
\volume{000}
\pagerange{000--000}
\date{\today}
\begin{document}
\maketitle

\begin{abstract}
We study the flow induced by random vibration of a solid
boundary in an otherwise quiescent fluid. The analysis is motivated by
experiments conducted under the low level and random effective
acceleration field that is 
typical of a microgravity environment. When the boundary is planar and
is being vibrated along its own plane, the variance of the velocity field
decays as a power law of 
distance away from the boundary. If a low frequency cut-off is introduced in
the power spectrum of the boundary velocity, the variance
decays exponentially for distances larger than a Stokes layer thickness
based on the cut-off frequency. Vibration of a gently curved boundary
results in steady streaming in the ensemble average of the tangential
velocity. Its amplitude diverges logarithmically with distance away
from the boundary, but asymptotes to a constant value instead
if a low frequency cut-off is considered. This steady component of the 
velocity is shown to depend
logarithmically on the cut-off frequency. Finally, we consider the case
of a periodically modulated solid boundary that is being randomly vibrated.
We find steady streaming in the ensemble average of the first order velocity,
with flow extending up to a characteristic distance of the order of the 
boundary wavelength. The structure of the flow in the vicinity of the 
boundary depends strongly on the correlation time of the boundary velocity.
\end{abstract}

\section{Introduction}
\label{sec:introduction}

This paper examines the formation and separation of viscous layers 
in a fluid which is in contact with a
solid boundary that is vibrated randomly. The analysis is motivated by the low
level and random acceleration field that affects a number of microgravity
experiments. We first study the case of a planar boundary to
generalize the classical result of \cite{re:stokes51} who considered
a boundary vibrated periodically along its own plane.
We next consider a slightly curved boundary, and show that steady streaming 
appears in the ensemble average at first order in the perturbed flow
variables. There are
several qualitative similarities and differences with the classical result by
\cite{re:schlichting79} for the case of periodic vibration. Finally, we
address the case of a modulated boundary that is vibrated randomly.

Our study is motivated by the significant levels of residual accelerations
(g-jitter) that have been detected during space missions in which microgravity 
experiments have been conducted
(\cite{re:walter87,re:nelson91,re:delombard97}). 
Direct measurement of these residual accelerations has shown that they have a 
wide frequency spectrum, ranging approximately from $10^{-4} Hz$ to $10^{2}
Hz$. Amplitudes range from  $10^{-6}g_{E}$ at the lowest end of
the frequency spectrum, and increase roughly linearly for high
frequencies, reaching values of $10^{-4}g_{E}\;-\;10^{-3}g_{E}$
at frequencies of around $10 Hz$ ($g_{E}$ is the intensity of
the gravitational field on the Earth's surface). Despite the efforts of a 
number of researchers over the last decade, there remain areas of uncertainty
about the potential effect of such a residual acceleration field on
typical microgravity fluid experiments, especially in quantitative terms. 
A better understanding of the response of a fluid to such 
disturbances would enable improved experiment design to minimize 
or compensate for their influence. In addition, it would also be useful to 
have error estimates of quantities measured in the presence of residual 
accelerations, including whenever possible some methodology for extrapolation 
to ideal zero gravity.

The formation of viscous layers around solid boundaries 
when the flow amplitude has a random component has not been addressed yet
despite its potential relevance for a number of microgravity experiments. 
Among them we
mention the dynamics of colloidal suspensions, coarsening studies of 
solid-liquid mixtures in which purely diffusive controlled transport
is desired,
or the interaction between the viscous layer produced by bulk flow of random 
amplitude and the morphological instability of a crystal-melt interface.
Our study represents the first step in this direction, and focuses on simple
geometries in order to elucidate those salient features of the flow that arise 
from the random nature of the vibration.

Previous theoretical work on the influence of g-jitter on fluid flow
ranges from order of magnitude estimates to
detailed numerical calculations. For example, the order of magnitude of
the contributions to fluid flow from the residual acceleration field
may be estimated from the length and time scales of a particular
experiment, and the values of the relevant set of dimensionless numbers
(\cite{re:alexander90}). Such studies are of interest as a first approximation,
but are not very accurate. Other studies have modeled the residual
acceleration field by some simple analytic function in which the acceleration 
is typically decomposed into steady and time dependent components, the latter 
being periodic in time
(\cite{re:gershuni76,re:kamotani81,re:alexander91,re:farooq94,re:grassia98a,re:grassia98b,
re:gershuni98}).
A few studies have also addressed the consequences of isolated pulses
of short duration (\cite{re:alexander97}).

\cite{re:zhang93} and \cite{re:thomson97} adopted a statistical 
description of the residual acceleration field onboard spacecraft, and modeled 
the acceleration time series as a stochastic process in time.
The main premise of this approach is that a statistical
description is necessary in those cases in which the characteristic
time scales of the physical process under investigation are long
compared with the correlation time of $g$-jitter, $\tau$ (the acceleration
amplitudes and orientations at two different times are statistically 
independent if separated by an interval larger than $\tau$).
Progress has been achieved through the consideration of a specific stochastic 
model according to which each Cartesian component of the residual acceleration 
field $\vec{g}(t)$ is modeled as a narrow band noise. This noise is a Gaussian 
process defined by three independent parameters: its intensity 
$\left<g^2\right>$, a dominant angular frequency $\Omega$, and a 
characteristic spectral width $\tau^{-1}$. Each realization of narrow band 
noise can be viewed as a temporal sequence of periodic functions of angular 
frequency $\Omega$ with amplitude and phase that remain constant only for a 
finite amount of time ($\tau$ on average). At random intervals, new values of 
the amplitude and phase are drawn from prescribed distributions. This model is 
based on the following mechanism underlying the residual acceleration field: 
one particular natural frequency of vibration of the spacecraft structure 
($\Omega$) is excited by some mechanical disturbance inside the spacecraft, 
the excitation being of random amplitude and taking place at a sequence of 
unknown (and essentially random) instants of time. 

Narrow band noise has been 
shown to describe reasonably well many of the features of g-jitter time 
series measured onboard Space Shuttle by \cite{re:thomson97}. Actual 
$g$-jitter data collected during the SL-J mission were analyzed, and a time 
series of roughly six hours was studied in detail. A scaling analysis revealed 
the existence of both deterministic and stochastic components in the time 
series. The deterministic contribution appeared at a frequency of $17$ Hz, 
with an amplitude $\left<g^2\right>^{1/2}= 3.56 \times 10^{-4} g_{E}$. 
Stochastic components included two well defined spectral features with a 
finite correlation time; one at 22 Hz with $\left<g^2\right>^{1/2}= 3.06 
\times 10^{-4}g_{E}$ and $\tau = 1.09$ s, and one at 44 Hz with 
$\left<g^2\right>^{1/2}= 5.20 \times 10^{-4} g_{E}$ and $\tau = 0.91$ s. White 
noise background is also present in the series with an intensity
$D = 8.61 \times 10^{-4} {\rm cm}^{2}/{\rm s}^{3}$.

A further theoretical advantage of narrow band noise is that it provides a 
convenient way of
interpolating between monochromatic noise (akin to studies involving a
deterministic and periodic gravitational field), and white noise (in which
no frequency component is preferred). In the limit $ \tau\rightarrow 0$ with
$D = \left< g^{2}\right> \tau$ finite, narrow band noise reduces to white noise
of intensity $D$; whereas, for $\tau\rightarrow\infty$ with $\left< g^{2}
\right>$ finite, monochromatic noise is recovered. 

We discuss in this paper the flow induced in an otherwise quiescent fluid by 
the random vibration of a solid boundary. The velocity of the boundary 
$u_{0}(t)$ is assumed prescribed, and modeled as a narrow band stochastic 
process. First, we consider an infinite planar boundary that is being vibrated 
along its own plane to generalize the classical problem studied 
by \cite{re:stokes51}. In the monochromatic limit, the variance of the
velocity field decays exponentially away from the wall, with a
characteristic decay
length given by the Stokes layer thickness $\delta_s=(2\nu/\Om)^{1/2}$,
where $\nu$ is the kinematic viscosity of the fluid, and $\Om$ is the angular
frequency of vibration of the boundary. Since the equations governing the flow
are linear, we are able to obtain an analytic solution describing transient
layer formation in the stochastic case, but only in the neighborhood of the 
white and monochromatic noise limits. We then show that for any finite 
correlation time the stationary variance of the tangential velocity 
asymptotically decays as 
the inverse squared distance from the wall, in contrast with the exponential
decay in the deterministic case. This asymptotic behavior originates from
the low frequency range of the power spectrum of the boundary velocity. The
crossover from power law to exponential decay is explicitly computed by
introducing a low frequency cut-off in the power spectrum.

We next investigate two additional geometries in which the equations governing
fluid flow are not linear, and show that several of the generic features
obtained for the case of a
planar boundary still hold. In the first case, we generalize the analysis of
\cite{re:schlichting79} concerning secondary steady streaming.
He found that the oscillatory motion of the boundary
induces a steady secondary flow outside of the viscous boundary layer 
even when the velocity of the boundary averages to zero. If the thickness of 
the Stokes layer, $\delta_s$, and the amplitude of oscillation, $a$, are small
compared with a characteristic length scale of the  boundary $L$ 
($\delta_s \ll L$, $a\ll L$), then the generation of secondary steady 
streaming may be described as follows. Vibration of the rigid boundary gives 
rise to an oscillatory and nonuniform motion of the fluid. The flow is
potential in the bulk, and rotational in the boundary layer because of
no-slip conditions 
on the boundary. The bulk flow applies pressure at the outer edge of boundary 
layer, which does not vary across the layer. The non uniformity
of the flow leads to vorticity convection in the boundary layer through 
nonlinear terms. Both convection and the applied pressure drive
vorticity diffusion, and thus induce secondary steady motion which does not 
vanish outside of the boundary layer. In the simplest case in which the far 
field velocity is a standing wave $U(x,t)=U(x)\ucos(\Om t)$, the
tangential component of the secondary steady velocity is,
\begin{equation}
u^{(s)}=-\frac{3}{4\Om}U\frac{\ud U}{\ud x},
\label{sch0}
\end{equation}
where $x$ is a curvilinear coordinate along the boundary. In fact, Eq. 
(\ref{sch0}) serves as the boundary condition for the stationary part of the 
flow in the bulk. Similar conclusions have been later reached by 
\cite{re:batchelor67} who studied sinusoidal oscillations of
nonuniform phase, and by \cite{re:gershuni98} who studied
monochromatic oscillations of a general form.

The second geometry that we address is the so called wavy wall 
(\cite{re:lyne71}). The deterministic limit in which a wavy boundary is being 
periodically vibrated has been studied by a number of authors, mainly to
address the interaction between the flow above the sea bed and ripple
patterns on it (\cite{re:lyne71,re:kaneko79,re:vittori89,re:blondeaux94} and 
references therein). \cite{re:lyne71} deduced the existence of steady
streaming in the limit in which the amplitude of the wall deviation
from planarity is small compared with the thickness of the Stokes
layer. He introduced a conformal transformation and obtained an explicit
solution in the limit of small $k\Rey$, where $k$ is the wavenumber of
the wall profile scaled by the thickness of the Stokes layer, and 
$\Rey$ is the Reynolds number. The detailed structure of the secondary
flow depends on the ratio between the wavelength of the boundary
profile and the thickness of the Stokes layer.

In Sections \ref{sec:sch} and \ref{sec:wavy_wall}, we discuss how the results 
for these two geometries generalize to the case of stochastic vibration.
Section \ref{sec:sch} addresses the flow created by a gently curved solid
boundary that is being vibrated randomly. The perturbation parameter that we use
is the ratio between the amplitude of vibration and the characteristic inverse curvature
of the wall. The ensemble average of the stream function is not zero, 
and hence there exists stationary streaming in the
stochastic case as well. The average velocity diverges logarithmically away 
from the boundary because of the low frequency range of the power spectrum. 
We again introduce a low frequency cut-off $\om_c$ in the spectrum, and study the
dependence of the stationary streaming on the cut-off frequency. 
We compute the stationary tangential velocity as a function of $\om_c \ll 1$
and arbitrary $\beta$, and find a weak (logarithmic) singularity as 
$\om_c \rightarrow 0$.

Section \ref{sec:wavy_wall} discusses the formation of a boundary layer around
a wavy boundary that is vibrated randomly. Positive and negative vorticity
production in adjacent regions of the boundary introduces a natural
decay length
in the solution, thus leading to exponential decay of the flow away from the
boundary, even in the absence of a low frequency cut-off in the power spectrum
of the boundary velocity. Steady streaming is found at second order comprising
two or four recirculating cells per period of the boundary profile. The number
of cells depends on the scaled correlation time $\Om \tau$.

\section{Randomly vibrating planar boundary}
\label{sec:plane}

We first examine the case of a planar boundary that is being vibrated along 
its own plane. In this case the governing equations are considerably simpler 
then in the more general geometries discussed in Sections \ref{sec:sch} 
and \ref{sec:wavy_wall}. In particular, the Navier-Stokes equation is linear, 
fact that allows a complete solution of the flow. Nevertheless, this simple 
solution still exhibits several of the qualitative features that are present in
the case of random forcing by a curved boundary, namely asymptotic power law 
decay of the velocity field away from the boundary, and sensitive dependence 
on the low frequency range of the power spectrum of the boundary velocity.

Consider an infinite solid boundary located at $z=0$, and an
incompressible fluid that 
occupies the region $z > 0$. The Navier-Stokes equation, and boundary 
conditions are,
\begin{equation}
  \partial_t{u} = \nu\partial_z^2{u},
\label{plane1}
\end{equation}
\begin{equation}
  u(0,t)=u_0(t), \quad u(\infty,t) < \infty,
\label{plane2}
\end{equation}
where $z$ is the coordinate normal to the boundary, $u(z,t)$ is the $x$
component of the velocity, and $u_{0}(t)$ is the prescribed velocity of the 
boundary. The solution for harmonic vibration $u_0(t)=u_0\ucos(\Om t)$
was given
by \cite{re:stokes51}. It is a transversal wave that propagates into the bulk
fluid with an exponentially decaying amplitude,
\begin{equation}
u(z,t)=u_0\ue^{-z/\delta_s}\ucos(\Om t - z/\delta_s),
\label{stokes}
\end{equation}
where $\delta_s=(2\nu/\Om)^{1/2}$ is the Stokes layer thickness.

\subsection{Narrow band noise}

As discussed in the introduction, the main topic of this paper is to 
examine how the nature of the bulk flow changes when the boundary velocity 
$u_{0}(t)$ is a random process. Specifically, we consider a Gaussian process 
defined by
\begin{equation}
  \left< u_0(t)\right>=0, \quad
  \left<u_{0}(t)u_{0}(t')\right>=\left<{u_0}^{2}\right>
  \ue^{-|t-t'|/\tau}\ucos\Om(t-t').
\label{nbn2}
\end{equation}
This process is known as narrow band noise (\cite{re:stratonovich67}).
It is defined by three independent parameters: its variance 
$\left<u_0^2\right>$, its dominant angular frequency $\Om$, and the 
correlation time $\tau$. Each realization of this random process can be viewed 
as a sequence periodic functions of frequency $\Om$, with amplitude and phase 
that remain constant for a time interval $\tau$ on average. White noise is 
recovered when $\Om\tau\rightarrow 0$ while $D = \left<{u_0}^{2}\right>\tau$
remains finite, whereas the monochromatic noise limit corresponds to
$\Om\tau\rightarrow \infty$, with $\left<{u_0}^{2}\right>$ finite.
Monochromatic noise is akin to a single frequency periodic signal of the same
frequency, but with randomly drawn amplitude and phase. The relationship 
between the two can be illustrated by considering a deterministic function 
$x(t)=x_0\ucos(\Om t)$ and defining the temporal average as,
\begin{equation}
  \left<x(t)x(t')\right> =
  \lim_{T\rightarrow\infty}\frac{1}{T}\int_0^T\ud t\,x(t)x(t')=
  \frac{x_0^2}{2}\ucos(\Om(t-t')).
  \label{nbn21}
\end{equation}
This average coincides with the ensemble average of the noise when 
$\left<u_0^2\right>=x_0^2/2$. The power spectrum corresponding to the 
correlation function (\ref{nbn2}) is
\begin{equation}
  P(\omega)=\frac{\left<u_0^2\right>}{2\upi}\left[\frac{\tau}
  {1+\tau^2(\omega-\Omega)^2}+\frac{\tau}{1+\tau^2(\omega+\Omega)^2}\right].
  \label{nbn3}
\end{equation}
We will also use the spectral density of the process $u_0(t)$,
\begin{equation}
  \hat{u}_0(\om)=\frac{1}{2\upi}\int_{-\infty}^{\infty}
  \ud t\,u_0(t)\ue^{-\ui\om t},
  \label{spec_dens}
\end{equation}
so that its ensemble average and correlation function are respectively given by
\refstepcounter{equation}
$$
  \left<\hat{u}_0(\om)\right>=0,\quad
  \left<\hat{u}_0(\om)\hat{u}^{*}_{0}(\om')\right>=\delta(\om-\om')P(\om).
  \eqno{(\theequation{\mathit{a},\mathit{b}})}
  \label{spec_dens_cor}
$$
We will often use dimensionless variables in which $\left<u_0^2\right>/\Om$ 
is the scale of $P(\om)$, and $\Om$ is the angular frequency scale. In
dimensionless form,
\begin{equation}
  P(\omega,\beta)=\frac{1}{2\upi}\left[\frac{\beta}{1+\beta^2(\omega-1)^2}+
  \frac{\beta}{1+\beta^2(\omega+1)^2}\right],
  \label{nbn6}
\end{equation}
where $\beta=\Omega\tau$.
We have $\int_{-\infty}^{\infty}\ud\om P(\om,\beta)=1$, independent of $\beta$,
and also $\lim_{\beta \rightarrow \infty} P(\omega)
= \left[\delta(\omega-1)+\delta(\omega+1)\right]/2$.
Note that the power spectrum does not vanish at small frequencies. Instead,
$P(0,\beta)=\beta/(\upi(1+\beta^2)$, which for large and small $\beta$
behaves as $P(0,\beta)\sim 1/(\upi \beta)$ and $P(0,\beta)\sim \beta/\upi$
respectively. We will discuss separately the effect of this
low frequency contribution on the results presented in the remainder of the
paper.

\subsection{Transient layer formation}
\label{subsec:transient}

In the two limiting cases of white and monochromatic noise, it is possible
to find an analytic solution for the transient flow starting from an initially
quiescent fluid. The solution can be found, for example, by introducing
the retarded, infinite space Green's function  corresponding to
equation (\ref{plane1}), with boundary conditions (\ref{plane2}),
\begin{equation}
  G(z,t|z',t')=\frac{1}{(4\upi\nu(t-t'))^{1/2}}\left[
  \ue^{-(z-z')^2/4\nu(t-t')}-\ue^{-(z+z')^2/4\nu(t-t')}\right],
  \quad t>t',
\label{greenplane}
\end{equation}
and $G(z,t|z',t')=0$ for $t<t'$. If the fluid is initially quiescent,
$u(z,0)=0$, we find
\begin{equation}
u(z,t)=\nu\int_{0}^{t}\ud t'\,u_0(t')\left(\p_{z'}G\right)_{z'=0},
\label{quis_sol}
\end{equation}
with
\begin{equation}
  \left(\p_{z'}G\right)_{z'=0}=\frac{z}{[4\upi\nu^3(t-t')^3]^{1/2}}\,
  \ue^{-z^2/4\nu(t-t')}.
\label{dif_green}
\end{equation}
Equations (\ref{quis_sol}-\ref{dif_green}) determine the transient behavior for
any given $u_0(t)$. 

If $u_0(t)$ is a Gaussian, white noise process, the ensemble average of 
Eq.(\ref{quis_sol}) yields $\left<u(z,t)\right>=0$. The corresponding equation 
for the variance reads
\begin{equation}
\left<u^2(z,t)\right>=2D\nu^2\int_0^t\ud t'\,\left[\left(\p_{z'}G\right)_{z'=0}
\right]^2=
\frac{2D\nu}{\upi z^2}\left(1+\frac{z^2}{2\nu t}\right)\,\ue^{-z^2/2\nu t}.
\label{wn_var}
\end{equation}
The variance of the induced fluid velocity propagates into the fluid 
diffusively.
Saturation occurs for $t\gg z^2/2\nu$, at which point the variance does 
not decay exponentially far away from the wall, but rather as a power law.
\begin{equation}
\left<u^2(z,\infty)\right>=\frac{2D\nu}{\upi z^2}.
\label{wn_var_longtime}
\end{equation}
Ascertaining whether random vibration can induce flows far away from
the boundary in more general geometries is one of the main 
motivations for this paper.

Consider now the opposite limit of monochromatic noise with
correlation function
\begin{equation}
\left<u_0(t)u_0(t')\right>=\left<u_0^2\right>\ucos\left[\Om(t-t')\right] .
\label{monocf}
\end{equation}
Now using Eqs. (\ref{quis_sol}) and (\ref{monocf}) we find, 
(\cite{re:carslaw59}),
\begin{equation}
\frac{\left<u^2(z,t)\right>}{2\left<u_0^2\right>}=\frac{2}{\upi}
\int_{\kappa}^{\infty}\ud\sigma\,\ue^{-\sigma^2}
\int_{\kappa}^{\infty}\ud\mu\,\ue^{-\mu^2}
\ucos\left[\frac{z^2}{2\delta_s^2}\left(\frac{1}{\mu^2}-\frac{1}{\sigma^2}
\right)\right],
\label{monovar}
\end{equation}
with $\kappa = z/(4\nu t)^{1/2}$. A closed form solution can only be obtained 
for long times. We find,
\begin{equation}
\frac{\left<u^2(z,t)\right>}{2\left<u_0^2\right>}=\frac{\ue^{-2z/\delta_s}}{2}
+\frac{2\kappa^3\delta_s^2}{\upi^{1/2}z^2}\ue^{-z/\delta_s}\usin(\Om\,t-z/
\delta_s)+{\cal O}(\kappa^5(t)).
\label{monovarcor}
\end{equation}
At long times,  the variance propagates into the bulk with phase velocity 
$(2\nu\Om)^{1/2}$, while its amplitude decays exponentially in space over the 
scale of the Stokes layer, and as $t^{-3/2}$ in time. In summary, the flow
created by the vibration of the boundary propagates diffusively for 
white noise $(z^2\propto 2\nu t)$, and as a power law 
$(z^2\propto \upi\nu\Om^2 t^3)$ in the monochromatic limit.

\subsection{Stationary variance for narrow band noise}

We have been unable to obtain a closed analytic solution for the transient 
evolution of the variance $\left< u(z,t)^{2} \right>$ when the
vibration of the 
boundary is given by a general narrow band process. It is possible, 
however, to obtain the stationary variance of the velocity. Equation 
(\ref{plane1}) can be rewritten in Fourier space as
\begin{equation}
\ui\om\hat{u}(z,\om)=\nu\p^2_z\hat{u}(z,\om)
\label{fourier2}
\end{equation}
with
\begin{equation}
u(z,t)=\int_{-\infty}^{\infty}\ud\om\,\hat{u}(z,\om)\ue^{\ui\om t},
\label{fourier1}
\end{equation}
The boundary conditions are,
$\hat{u}(0,\om)=\hat{u}_0(\om)$, and $\hat{u}(z,\om)<\infty$ at $z \rightarrow
\infty$. The solution of Eq. (\ref{fourier2}) with these boundary conditions is
\begin{equation}
  \hat{u}(z,\om)=\ue^{-\alpha z}\hat{u}_0(\om),
  \quad \alpha=(1+\ui\,\usign(\om))(\om/2\nu)^{1/2}
\label{fourier4}
\end{equation}

We next choose $1/\Om$ as the time scale, and $\delta_s$ as the length scale, 
and after some straightforward algebra we find,
\begin{equation}
  \frac{\left<u^2(z,\beta)\right>}{2 \left<u_{0}^{2}\right>} = I(z,\beta)=
\int_{0}^{\infty} \ud\om P(\om,\beta)e^{-z\om^{1/2}}.
\label{plane3}
\end{equation}
We have also used the fact that the power spectrum (\ref{nbn6}) is even 
in frequency.

We next analyze the asymptotic dependence of $I(z,\beta)$ at large $z$.
In this limit, $I(z,\beta)$ mainly depends on the low frequency region of the
power spectrum; higher frequencies are suppressed by the exponential factor.
By using Watson's lemma (\cite{re:nayfeh81}), we find,
\begin{equation}
I(z,\beta)=\frac{2\beta}{\upi(1+\beta^2)}\frac{1}{z^2} +
\frac{240\beta^3(3\beta^2-1)}{\upi(1+\beta^2)^3}\frac{1}{z^6} +
{\cal O}(z^{-10}).
\label{plane4}
\end{equation}
This asymptotic form at large $z$ is valid for all $\beta$.
In particular, the dominant behavior for small and large $\beta$ is
\begin{displaymath}
I(z,\beta) \sim \frac{2\beta}{\upi}\frac{1}{z^2},\quad
I(z,\beta) \sim \frac{2}{\upi\beta}\frac{1}{z^2},
\end{displaymath}
respectively. Hence we recover the power law decay of Eq. 
(\ref{wn_var_longtime}).

\begin{figure}
  \vspace{1.5pc}
  \centerline{\epsffile{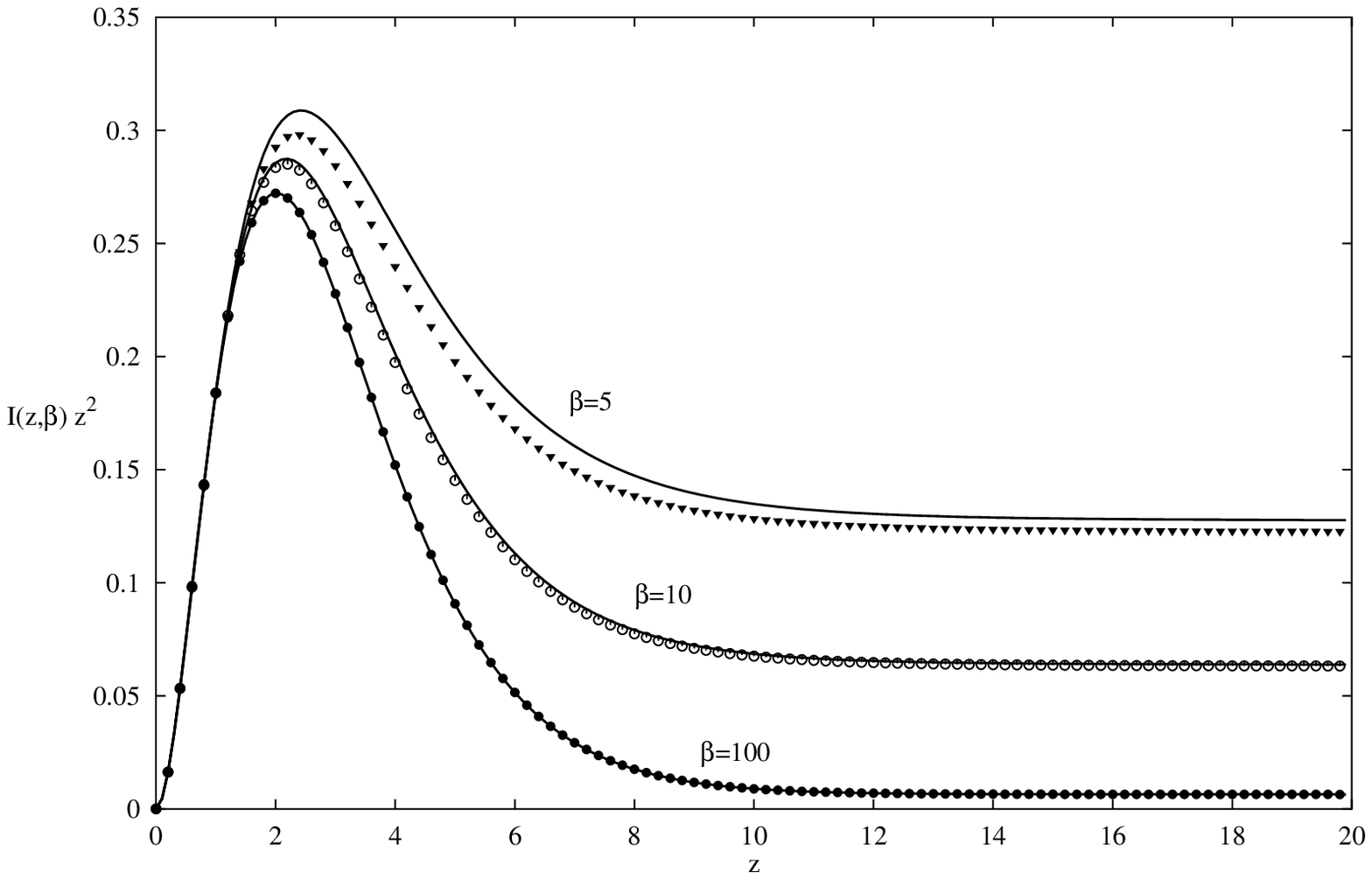}}
  \caption{Normalized variance of the tangential velocity for the case
of a planar boundary computed by numerical integration of Eq. (\ref{plane3})
(symbols), and its uniform asymptotic expansion, Eq. (\ref{plane5}), 
(solid lines). The function $I(z,\beta)\,z^2$ asymptotes to a constant
value outside of the classical Sokes layer based on $\Om$. The uniform expansion remains 
a good approximation even for moderate $\beta$.}
  \label{fig:plane_var}
\end{figure}

We can also find the asymptotic behavior at large $\beta$ that is 
uniformly valid in $z$,
\begin{equation}
I(z,\beta)=\frac{\ue^{-z}}{2}-\frac{z}{2\upi\beta}\left(
\Ci(z)\sin(z)-\Si(z)\cos(z)-\frac{1}{2}(\ue^{-z}\Ei(z)-\ue^z\Ei(-z))
\right) + {\cal O}(\beta^{-3}),
\label{plane5}
\end{equation}
where $\Ci$ and $\Si$ denote the integral sine and cosine functions, 
and $\Ei$ stands for the exponential integral function 
(\cite{re:gradshteyn80}).
For $z\lesssim 1$, the variance decreases exponentially. At larger $z$,
the exponential terms in (\ref{plane5}) become small, so that the remaining 
asymptotic dependence for large $z$ is given by Eq. (\ref{plane4}).
The quantity $I(z,\beta)\,z^2$ computed both from (\ref{plane3}) and the
uniform expansion (\ref{plane5}) is presented in Fig.\ref{fig:plane_var}.
For fixed $\beta$, $I(z,\beta)$ asymptotes to $2\beta/\upi(1+\beta^2)$
outside of Stokes layer. This value is the coefficient of the
leading term in the asymptotic expansion (\ref{plane4}). The expansion 
(\ref{plane5}) is a good approximation even for moderate values of $\beta$.

To summarize, the variance of the velocity field does not decay
exponentially away from the wall for finite $\beta$, but rather as the 
inverse squared distance. The crossover length separating exponential
and power law decay increases with increasing $\beta$.

\subsection{Low frequency cut-off in the power spectrum}
\label{subsec:cutplane}

The coefficient of the leading term in (\ref{plane4}) is in fact
twice the value of $P(0,\beta)=\beta/\upi(1+\beta^2)$. The algebraic decay of 
$\left<u_0^2(z,t)\right>$ follows from the diffusive nature of Eq. 
(\ref{plane1}), and a non vanishing value of $P(\om,\beta)$ as $\beta
\rightarrow 0$. Before we analyze in Sections \ref{sec:sch} and 
\ref{sec:wavy_wall} how this behavior is modified by nonlinearities in the
governing equations, we explicitly address here the consequences of a low 
frequency cut-off in the power spectrum. Of course, there always exists in 
practice a low frequency cut-off because of limited observation time. 
Furthermore, the low frequency range of the power spectrum of the residual 
acceleration field in microgravity ($\Om/2 \upi < 10^{-3}$Hz) is
fairly difficult to measure reliably. We therefore introduce an
effective cut-off frequency in the power spectrum, $\om_c\ll1$, and
study the dependence of $\left<u_0^2(z,t)\right>$ 
on $\om_c$. The stationary value of variance of the velocity is now given by,
\begin{equation}
  \frac{\left<u^2(z,\beta,\om_c)\right>}{2\left<u_0^2\right>}
  = I_c(z,\beta,\om_c)=\int_{\om_c}^{\infty}
  \ud\om P(\om,\beta)e^{-z\om^{1/2}}.
\label{plane1_cut}
\end{equation}
By using Watson's lemma, we find for large $z$,
\begin{equation}
 I_c(z,\beta,\om_c)=\frac{2\beta\ue^{-z\om_c^{1/2}}}{\upi(1+\beta^2)}
\left(\frac{1}{z^2}+\frac{\om_c^{1/2}}{z}+\uhot\right),
\label{plane2_cut}
\end{equation}
where $\uhot$ stands for terms which are of higher order than terms 
retained under the assumption that both $1/z$ and $\om_c^{1/2}$
are small but independent. For $z \gg 1$, but $z \om_c^{1/2} \ll 1$ 
the dominant term in (\ref{plane2_cut}) is 
\begin{equation}
  I_c(z,\beta,\om_c)\sim
  \frac{2\beta}{\upi(1+\beta^2)}\frac{1}{z^2},\quad z\om_c^{1/2}<<1.
\label{plane3_cut}
\end{equation}
On the other hand, if $z\om_c^{1/2}\geq 1$, the leading order term is now
a function of $\zeta=z\om_c^{1/2}$
\begin{equation}
  I_c(z,\beta,\om_c)\sim
  \frac{2\beta\om_c\ue^{-\zeta}}{\upi(1+\beta^2)}
  \left(\frac{1}{2\zeta}+\frac{1}{\zeta^2}\right),\quad \zeta\geq 1
\label{plane4_cut}
\end{equation}
Equations (\ref{plane3_cut}) and (\ref{plane4_cut}) show that at 
distances that are large compared with the thickness of the Stokes
layer based on the dominant frequency $\Om$, $\left<u^2(z,t)\right>$
decays algebraically with $z$. There exists, however, a length scale 
$z\om_{c}^{1/2}$ beyond which the decay is exponential. 
This new characteristic length scale is the thickness of the Stokes
layer based on the cut-off frequency. This conclusion appears natural 
given the principle of superposition for the linear differential
equation (\ref{plane1}).

\section{Streaming due to random vibration}
\label{sec:sch}

Next we investigate to what extent the results of Section \ref{sec:plane} hold
in configurations in which the governing equations are not linear. We examine 
in this section the flow induced by a gently curved solid boundary
that is being randomly vibrated. The boundary velocity is assumed to be
described by a narrow band stochastic process, and hence our results will
reduce to Schlichting's in the limit of infinite correlation time. 
However, for finite values of $\beta$ the results are qualitatively different.
The mechanism of secondary steady streaming generation described in the
introduction is no longer valid because there is no boundary layer solution at 
zeroth order. Vorticity produced at the vibrating boundary penetrates 
into the bulk fluid, at least perturbatively for small curvature. 
This results in a logarithmic divergence of the ensemble average of
the first order velocity with distance away from the wall. A cut-off 
analysis is also presented, and similarly to that of Section
\ref{subsec:cutplane}, it reveals the existence of an effective boundary
layer of thickness based on the cut-off frequency. We then find 
an expression analogous to Eq. (\ref{sch0}) as a function of $\beta$ and 
$\om_c$.

Define the following dimensionless quantities,
\begin{equation}
\left.
\begin{array}{l}
\displaystyle
z=\tilde{z}[(\nu/\Om)^{1/2}],\quad x=\tilde{x}[L],\quad
t=\tilde{t}[\Om^{-1}],\quad\psi=\tilde{\psi}[(2\left<u_0^2\right>
\nu/\Om )^{1/2}] , \\[16pt]
\displaystyle
\quad \eps=(2\left<u_0^2\right>)^{1/2}/\Om L,\quad \Rey_p=2\left<u_0^2\right>/
\Om\nu,
\quad\tilde{\Delta}=\p_{\tilde{z}}^2+\eps^2/\Rey_p\p_{\tilde{x}}^2
\end{array}
\right\}
\end{equation}
Assume now that the characteristic scale of the boundary $L$ is large so that
$\eps$ is a small quantity. If the Reynolds number $Re_{p}$ is assumed to
remain finite, both conditions imply $\nu / \Om L^{2} \ll 1$. We next write the
governing equations and boundary conditions in the frame of reference co-moving
with the solid boundary and obtain for a two dimensional geometry (tildes are
omitted),
\begin{equation}
  \partial_t{\Delta\psi}+\epsilon\frac{\p(\psi,\Delta\psi)}{\p(z,x)}=
  \Delta^2\psi
\label{sch1}
\end{equation}
\refstepcounter{equation}
$$
  \psi=0, \quad \p_z\psi=0 \quad \mbox{at}\quad y=0,\qquad
  \eqno{(\theequation{\mathit{a},\mathit{b}})}
  \label{sch2}
$$
\begin{equation}
  \p_z\psi=2^{-1/2}U(x)u_0(t) \quad \mbox{at}\quad z=\infty,
  \label{sch3}
\end{equation}
where $x,z$ are the tangential and normal coordinates along the boundary, 
and $\psi$ is the stream function $\bu=(\p_z\psi,-\p_x\psi)$. We have also 
used the notation $\p(a,b)/\p(z,x) = (\p_za)(\p_xb)-(\p_xa)(\p_zb)$ for the
nonlinear term. The far field boundary condition is a nonuniform and random
velocity field, of the order of  $\left<u_0^2\right>^{1/2}$, with a spatially 
nonuniform amplitude $U(x)$, and a stochastic modulation $u_0(t)$ which is a 
Gaussian stochastic process with zero mean, and narrow band power spectrum.
We first expand the stream function as a power series of $\eps$, 
$ \psi=\psi_0+\eps\psi_1+\ldots $ and solve (\ref{sch1}) order by
order. At order $\eps^0$ we obtain the following equation,
\begin{equation}
  (\p_t{\p_z^2}-\p_z^4)\psi_0=0
  \label{sch5}
\end{equation}
with boundary conditions,
\refstepcounter{equation}
$$
  \psi_0=0, \quad \p_z\psi=0 \quad \mbox{at}\quad z=0,\qquad
  \eqno{(\theequation{\mathit{a},\mathit{b}})}
  \label{sch6}
$$
\begin{equation}
  \p_z\psi=2^{-1/2}U(x)u_0(t) \quad \mbox{at}\quad z=\infty .
  \label{sch7}
\end{equation}
At this order, the equations effectively describe the flow induced
above a planar boundary with a far field velocity boundary condition
that is not uniform. The solution can be found by Fourier
transformation. We define,
\begin{eqnarray}
  \psi_0(x,z,t)=\int_{-\infty}^{\infty}\ud\om\,\hat{\psi}_0(x,z,\om)
\ue^{\ui\om t}.
  \label{sch8}
\end{eqnarray}
The transformed Eq. (\ref{sch5}) and the transformed boundary conditions 
(\ref{sch6}) allow separation of variables. We define,
$$
  \hat{\psi}_0(x,z,\om)=2^{-1/2}U(x)\hat{u}_0(\om)\hat{\zeta}_0(z,\om),
$$
so that Eq. (\ref{sch5}) leads to,
\begin{equation}
  (\ui\om{\p_z^2}-\p_z^4)\hat{\zeta}_0=0,
  \label{sch51}
\end{equation}
with boundary conditions
$\hat{\zeta}_0=0$, $\p_z\hat{\zeta}_0=0$, at $z=0$ and
$\p_z\hat{\zeta}_0=1$ at $z=\infty$. The solution is,
\begin{equation}
       \hat{\zeta}_0(z,\om)=-1/\alpha+z+1/\alpha\ue^{-\alpha z},\quad
       \alpha(\om) = (1+\ui\,\usign(\om))(|\om|/2)^{1/2}).
\label{sch81}
\end{equation}
 
At order $\eps$ we find,
\begin{equation}
  (\p_t{\p_z^2}-\p_z^4)\psi_1=\p_x\psi_0\p_z^3\psi_0 - 
  \p_z\psi_0\p_x\p_z^2\psi_0
  \label{sch9}
\end{equation}
with boundary conditions $ \psi_1=0, \; \p_z\psi_1=0$ at $y=0$.
The remaining boundary condition for $\psi_1$ needs to be discussed separately.
Consider first the classical deterministic limit which can be 
formally obtained by setting $\beta=\infty$. Then, the right hand side of 
Eq. (\ref{sch9}) involves stationary terms (of zero frequency), and sinusoidal 
terms with twice the frequency of the far field flow. Since the equation is 
linear, the solution $\psi_1$ has exactly the
same temporal behavior. In this case, it is known that it is not possible to 
find a solution for $\p_z\psi_{1}$ that vanishes at large $z$, but
only one that simply remains bounded as $z \rightarrow \infty$.  By
analogy, we introduce a similar requirement in the stochastic case of 
$\beta < \infty$. Since the
zeroth order stream function diverges linearly, this condition simply
amounts to requiring that the expansion in powers of $\eps$ remains consistent.

We now take the ensemble average of Eq. (\ref{sch9}), and consider the long
time stationary limit of the average 
($\psi_1^{(s)}=\lim_{t\rightarrow\infty}\left<\psi_1\right>$), to find,
\begin{equation}
  \p_z^4\psi_1^{(s)}=\left<\p_z\psi_0\p_x\p_z^2\psi_0-\p_x\psi_0\p_z^3\psi_0
  \right>.
  \label{sch101}
\end{equation}
The right hand side of this equation can be integrated from infinity 
to $z$. We obtain,
\begin{equation}
  \p_z^3\psi_1^{(s)}=\frac{U}{2}\frac{\ud U}{\ud x}F(z,\beta),
  \label{sch11}
\end{equation}
where
$$
F(z,\beta)=\int_{0}^{\infty}\ud\om P(\om,\beta)Q(\om,z),
$$
and
$$
  Q(\om,z)=  (-2+2\hat{\zeta}_0\p_z\hat{\zeta}_0^* -
	     \hat{\zeta}_0\p_z^2\hat{\zeta}_0^* -
	     \hat{\zeta}_0^*\p_z^2\hat{\zeta}_0).
$$
The power spectrum $P(\om,\beta)$ is defined in Eq. (\ref{nbn6}). The constant 
that appears in the expression for $Q(\om,z)$ comes from the pressure gradient 
imposed at infinity.

We now proceed to solve Eq.(\ref{sch11}) subject to the boundary conditions
$ \psi_1^{(s)}=\p_z\psi_1^{(s)}=0$ on the solid boundary, and 
$\p_z\psi_1^{(s)}<\infty$ as $z\rightarrow \infty$.
This is a boundary value problem for $\psi_1^{(s)}$. 
In the limit $\beta \rightarrow \infty$, it can be solved 
analytically, and the result obtained by Schlichting is recovered, namely that 
the solution may be bounded at infinity simply by setting the homogeneous 
part of the solution equal to zero to satisfy the principle of minimal 
singularity (\cite{re:vandyke64}). Otherwise $\p_z\psi_1^{(s)}$ is 
singular a fortiori. Since we cannot find an complete analytic
solution for finite $\beta$, we proceed as follows. We recast the 
boundary value problem as an initial value 
problem, and search for a boundary condition on $\p_z^2\psi_1^{(s)}$ at $z=0$ 
so that the homogeneous part of the solution remains bounded. This boundary 
condition can be found analytically by integrating (\ref{sch11}) from 
$0$ to $z$. We find,
$$
  \p_z^2\psi_1^{(s)}(z)-\p_z^2\psi_1^{(s)}(0)=
  \frac{U}{2}\frac{\ud U}{\ud x}\int_0^{z}\ud z'
  \int_{0}^{\infty}\ud\om P(\om,\beta)Q(\om,z'),
$$
or after changing the order of integration,
$$
\p_z^2\psi_1^{(s)}(z)-\p_z^2\psi_1^{(s)}(0)= 
\frac{U}{2}\frac{\ud U}{\ud x}
\int_{0}^{\infty}\ud\om P(\om,\beta)
\left[
\int_z \ud z' Q(\om,z')-\left(\int_z \ud z' Q(\om,z')\right)_{z=0}
\right].
$$
The second integral within brackets equals $(1/2\om)^{1/2}$. In order to 
avoid a linear divergence of $\p_z\psi_1^{(s)}(z)$ we equate the
constant terms on both sides, thus obtaining the third initial condition
\begin{eqnarray}
\p_z^2\psi_1^{(s)}(0,\beta) & = & 
U\frac{\ud U}{\ud x}
\frac{2^{1/2}}{4}\int_0^{\infty}\ud\om\,\om^{-1/2}P(\om,\beta) =
\nonumber\\
& = & U\frac{\ud U}{\ud x}\frac{(2\beta)^{1/2}}{8q}
((2(q-\beta))^{1/2}+(q+1)^{1/2}+(q-1)^{1/2}),
\label{sch12}
\end{eqnarray}
where $q = (1+\beta^2)^{1/2}$, and the dependence of initial condition 
on $\beta$ is shown explicitly. Equation (\ref{sch11}) with its
original boundary
conditions, supplemented with Eq. (\ref{sch12}) is an initial 
value problem, which we have solved numerically. 

Before presenting the numerical results, we study the
asymptotic behavior of the solution for large $z$ which
is determined by the asymptotic form of $F(z,\beta)$ at large $z$.
By explicit substitution of the zeroth order solution we find
\begin{equation}
Q(z,\om)=(-4\ucos(Z)-2Z(\ucos(Z)+\usin(Z))+2\usin(Z))\ue^{-Z}+2\ue^{-2Z},
\label{a2}
\end{equation}
where $Z = z(\om/2)^{1/2}$. The leading contribution to $F(z,\beta)$
as $z\rightarrow\infty$ originates from the zero frequency limit of
$P(\om,\beta)$. Thus $ F(z,\beta)\sim
P(0,\beta)\int_{0}^{\infty}\ud\om Q(\om,z)$.
The remaining integral may be easily calculated to yield the
asymptotic form,
\begin{equation}
F(z,\beta)\sim \frac{6\beta^2}{\upi(1+\beta^2)}\frac{1}{z^2} .
\label{sch13}
\end{equation}
Therefore the stationary mean
first order velocity $u_1^{(s)}=\p_z\psi_1^{(s)}$ has a logarithmic 
asymptotic form (see Eq. (\ref{sch11})). This is to be contrasted with 
the power law decay of the stationary variance for the case of a
randomly vibrating planar boundary.

\begin{figure}
  \vspace{1.5pc}
  \centerline{\epsffile{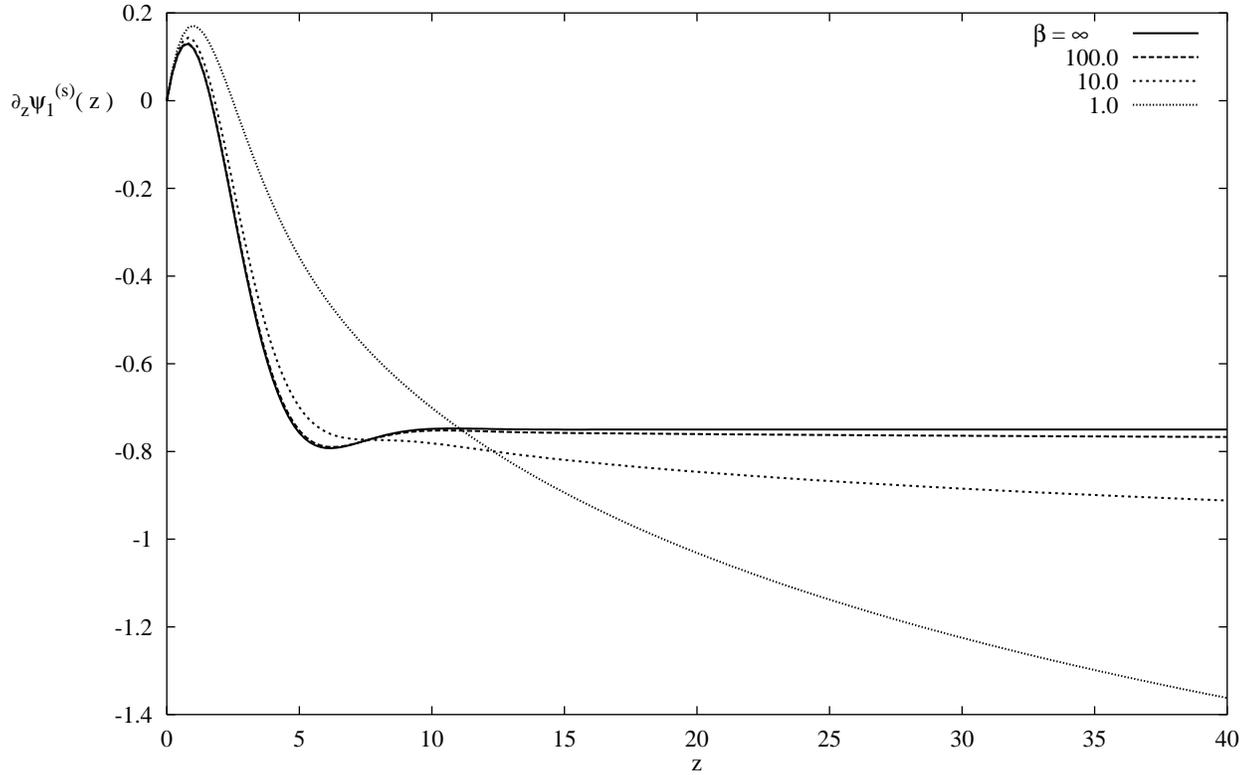}}
\caption{Stationary first order velocity as a function of distance for a
range of values of $\beta$. All curves diverge logarithmically at large $z$, 
except for $\beta=\infty$ (monochromatic limit), in which the velocity 
asymptotes to a constant within the Stokes layer. This latter behavior
reproduces the classical result of Schlichting.}
  \label{fig:sc_u_prof}
\end{figure}

Finally, we have numerically calculated $\partial_{z}\psi_{1}^{(s)}(z)$ for a 
range of values of $\beta$. The results are presented in Fig. 
\ref{fig:sc_u_prof}. Equation (\ref{sch11}) was integrated numerically
with no-slip boundary conditions for $\psi_1^{(s)}$, and Eq.
(\ref{sch12}). The numerical results support our conclusion
about the logarithmic divergence of $u_1^{(s)}$ with distance for any 
finite $\beta$. As $\beta$ increases the stationary mean velocity
profile approaches a limiting
form that corresponding to the monochromatic limit of 
$\beta \rightarrow \infty$.
In this limit we recover the Schlichting result, according to which 
$\partial_{z} \psi_{1}^{(s)}(z)$ asymptotes to a constant value over a 
distance of the order of the deterministic Stokes layer. 

\subsection{Low frequency cut-off in the power spectrum}
\label{subsec:cut_sch}

\begin{figure}
  \vspace{1.5pc}
  \centerline{\epsffile{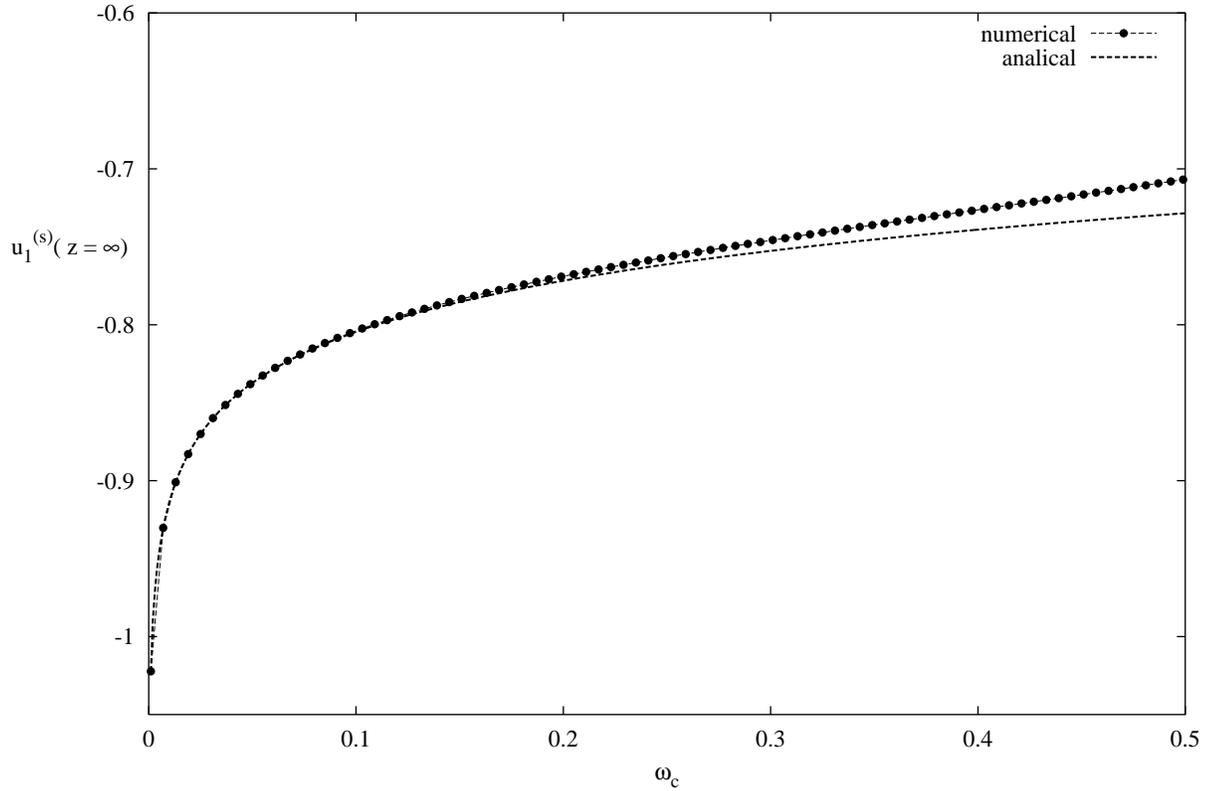}}
\caption{Asymptotic dependence of the stationary velocity as a function of 
the cut-off frequency $\om_c$. We show the case $\beta=10$ given by 
Eq. (\ref{sch_cut_asym2}) along with the numerically obtained solution.}
\label{fig:usc_cut}
\end{figure}

In Section \ref{sec:plane}, we showed that a low frequency cut-off in 
$P(\om,\beta)$ led 
to an exponential decay of the velocity outside an effective boundary
layer of thickness determined by the cut-off frequency. We therefore
examine here the consequences of a low frequency cut-off on the
divergent behavior of the stationary average of the first order stream
function. In order to find the asymptotic form of $\p_z\psi_1^{(s)}$, 
we first integrate (\ref{sch11}) twice 
from $z=0$ to $z$. By using the low frequency cut-off defined in Section 
(\ref{subsec:cutplane}), we write,
\begin{equation}
\p_z\psi_1^{(s)}(z,\beta,\om_c)=\frac{U}{2}\frac{\ud U}{\ud x}
\left[G_c(z,\beta,\om_c)-G_c(0,\beta,\om_c)\right] ,
\label{sch_cut_asym}
\end{equation}
with,
$$
G_c(z,\beta,\om_c)=\int \ud z'\int \ud z' F_c(z',\beta,\om_c),
$$
and,
$$
 F_c(z,\beta,\om_c)=\int_{\om_c}^{\infty}\ud\om P(\om,\beta)Q(\om,z).
$$
If we set $\om_c = 0$ and consider the monochromatic limit of $\beta 
\rightarrow \infty$, we find that $G_c(z,\infty,0)$ contains an exponential
factor that vanishes at $z\sim{\cal O}(1)$, and that $G_c(0,\infty,0) = 3/4$.
Therefore the Schlichting result is recovered. An explicit form of 
$G_c(z,\beta,\om_c)$ for any $\beta$ and $\om_c$ can be obtained
analytically, but it is far too complicated and we do not quote it here.
It has a similar functional dependence as in the cut-off analysis of
the planar boundary, and contains exponential terms involving
$(-z\om_c^{1/2})$. We find that $\p_z\psi_1^{(s)}(\infty,\beta,\om_c)=
-U\ud U/2\ud x\, G_c(0,\beta,\om_c).$ For finite but small $\om_c$, we find,
\begin{equation}
\p_z\psi_1^{(s)}(\infty,\beta,\om_c)=-\frac{3}{4}U\frac{\ud U}{\ud x}
\frac{\beta}{\upi(1+\beta^2)}
\left(2\beta\arctan(\beta)-\ln\left(\frac{\beta^2\om_c^2}{1+\beta^2}\right) 
\right) + {\cal O}(\om_c^2)
\label{sch_cut_asym2}
\end{equation}
This asymptotic formula is compared with the numerically computed
value of the tangential mean stationary velocity at large distances
in Figure \ref{fig:usc_cut}. Computations were done as described in the
previous section, and for different values of $\om_c\ll 1$ and
$\beta$. In all cases
$\p_z\psi_1^{(s)}$ reached constant values at large enough $z$ (the numerical
value of infinity, $z_{\infty}$, was chosen so that the change in velocity for
$z\geq z_{\infty}$ was less than a prescribed tolerance). We also checked
that any change in the boundary condition
Eq.(\ref{sch12}) leads to a linear divergence in the
tangential mean stationary velocity, thus confirming the adequacy of this 
boundary condition. The figure also shows that the computed values
of $\p_z\psi_1^{(s)}(z=\infty,\beta,\om_c)$ for small $\om_c$  are in a good
agreement with (\ref{sch_cut_asym2}).

In summary, Eq. (\ref{sch_cut_asym2}) shows that the tangential velocity away
from the boundary asymptotes to a constant that is a function of
$\beta$, and
has a weak (logarithmic) dependence on the cut-off frequency $\om_c$. 
Therefore the asymptotic dependence in the stochastic case (with a cut-off) 
and the deterministic case is
qualitatively similar, although the value of the asymptotic velocity of the
former depends on $\beta$. Note also that this asymptotic behavior only
sets in for distances larger than $(\nu/\om_{c}\Om)^{1/2}$, a value that can be
quite large in practical microgravity conditions.

\section{Randomly vibrating wavy boundary}
\label{sec:wavy_wall}

In the two previous sections we considered cases in which the characteristic
longitudinal length scale of the solid boundary was either infinite or large 
compared with
the characteristic amplitude of boundary vibrations. We examine here
the case of a wavy boundary and study how comparable length scales in
both directions influence the flow away from the boundary. In contrast
with the Schlichting problem, the external applied flow is now uniform
or, alternatively, the length scale over which the flow is not uniform
is much larger that the wavelength of the boundary. Thus one
anticipates that the normal component of the flow that appears
is caused by the wall profile. This flow interacts through
nonlinear terms with the externally forced flow that is parallel 
to the average profile of the boundary and gives rise
to stationary streaming. Even for stochastic vibration we show that positive 
and negative vorticity production in 
adjacent regions of the boundary introduces a natural decay length in the
zeroth order solution, thus leading to exponential decay of the flow
away from the boundary, even in the absence of a low frequency cut-off in the
power spectrum of the boundary velocity.

\begin{figure}
  \vspace{1.5pc}
  \centerline{\epsffile{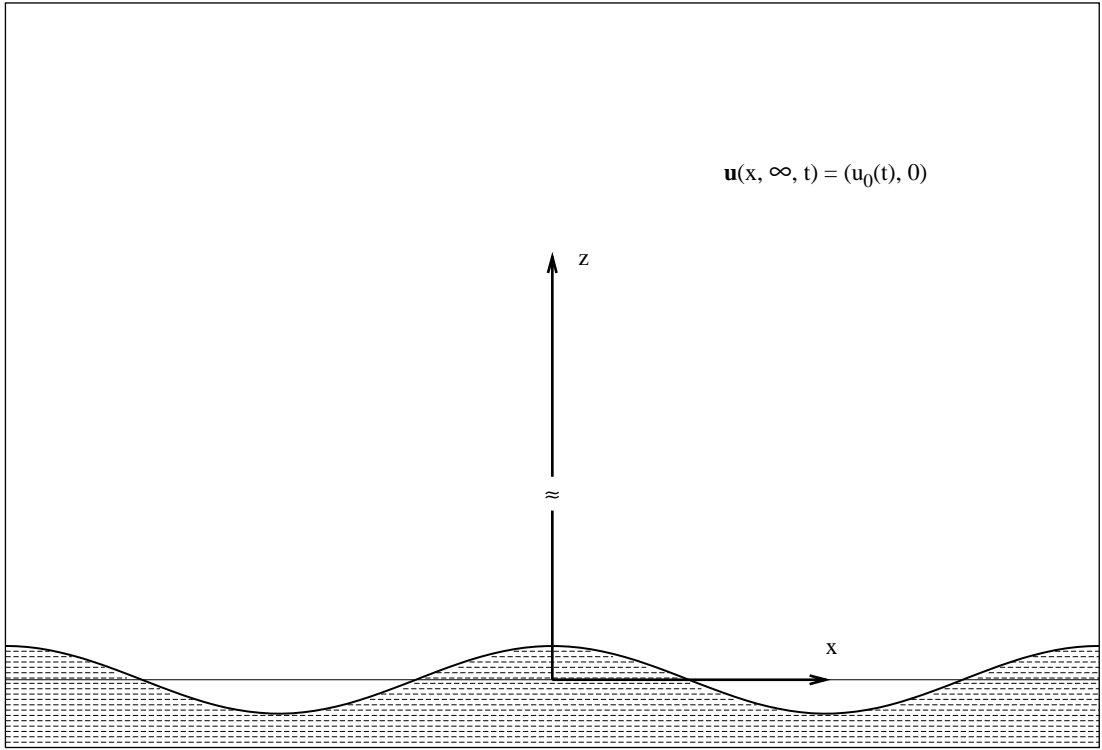}}
  \caption{Schematic view of the geometry of the wavy wall studied
  		in Section \ref{sec:wavy_wall}.}
  \label{fig:wavy_geom}
\end{figure}

Consider a rigid wavy wall being washed by a uniform oscillatory flow parallel 
to the wall wave vector,
\begin{equation}
\mathbf{u}(x,z=\infty)=(u_0(t),0).
\label{rvwb_uinf}
\end{equation}
We now assume that $u_0(t)$ is a narrow band Gaussian process.
Assume also that the amplitude of the boundary profile $l$ is small
compared with both the Stokes layer $\delta_s$ and the
wavelength $L$, with $\delta_{s}/L$ finite. The following dimensionless 
quantities are introduced,
\begin{equation}
\left.
\begin{array}{l}
\displaystyle
z=\tilde{z}[(\nu/\Om)^{1/2}],\quad x=\tilde{x}[(\nu/\Om)^{1/2}],\quad
t=\tilde{t}[\Om^{-1}],\quad\psi=\tilde{\psi}[(2\left<u_0^2\right> 
\nu/\Om )^{1/2}], \\[16pt]
\displaystyle
\eps=l/(\nu/\Om)^{1/2},\quad \Rey=\left[2\left<u_0^2\right>/
\Om\nu\right]^{1/2},
\quad k=2\upi(\nu/\Om)^{1/2}/L,\quad\tilde{\Delta}=\p_{\tilde{z}}^2+
\p_{\tilde{x}}^2
\end{array}
\right\}
\end{equation}
referred to the Cartesian coordinate system sketched in
Fig. \ref{fig:wavy_geom}. The solid boundary is located at
\begin{equation}
\tilde{\eta}(\tilde{x})=\hat{\eta}\epsilon\uexp(\ui k \tilde{x})+\ucc
\label{profile}
\end{equation}
with constant complex amplitude $\hat{\eta}$ so that $|\hat{\eta}|=1/2$. 
The dimensionless (and two dimensional) Navier-Stokes equation 
(tildes are omitted in what follows) reads,
\begin{equation}
  \partial_t{\Delta\psi}+\Rey\frac{\p(\psi,\Delta\psi)}{\p(z,x)}=
  \Delta^2\psi ,
\label{rvwb1}
\end{equation}
with no-slip conditions at the boundary,
\refstepcounter{equation}
$$
  \psi=0, \quad \p_z\psi=0 \quad \mbox{at}\quad y=\eta(x),\qquad
  \eqno{(\theequation{\mathit{a},\mathit{b}})}
  \label{rvwb2}
$$
and the imposed uniform flow at infinity,
\refstepcounter{equation}
$$
  \p_x\psi=0, \quad \p_z\psi=2^{-1/2}u_0(t) \quad \mbox{at}\quad z=\infty.
  \eqno{(\theequation{\mathit{a},\mathit{b}})}
  \label{rvwb3}
$$
These equations depend only on three dimensionless parameters: $\eps$,
the ratio of the amplitude of the wavy wall to the boundary layer
width; $\Rey$, the Reynolds number (the square root of $\Rey_p$ used in 
Section \ref{sec:sch}); and $k$, the wavenumber of the wall profile in units
of boundary layer width. We assume $\eps \ll 1$ and expand the
stream function in a power series of $\eps$,
\begin{equation}
  \psi=\psi_0+\eps\psi_1+\ldots
  \label{rvwb5}
\end{equation}
The boundary conditions are likewise expanded in power series of
$\eps$,
\begin{equation}
\psi(z)|_{z=\eta}=\psi_0(z)|_{z=0}+\eps(\psi_1(z)|_{z=0}+
\eta\p_z\psi_0(z)|_{z=0})+\ldots
\label{rvwb6}
\end{equation}
We now solve (\ref{rvwb1}) order by order in $\eps$. 

At zeroth order the wall is effectively planar. We decompose the Fourier
transform of the stream function as,
$ \hat{\psi}_0(x,z,\om)=2^{-1/2}\hat{u}_0(\om)\hat{\zeta}_0(z,\om)$.
The function $\hat{\zeta}_0(z,\om)$ is given in Eq. (\ref{sch81}). At
this order, the solution is identical to that found for a planar boundary.

At first order we seek a solution of the form,
\begin{equation}
  \psi_1=\hat{\eta} \; \uexp(\ui kx)\phi(z,t)+\ucc
\label{rvwb7}
\end{equation}
so that the amplitude $\phi(z,t)$ satisfies the Orr-Sommerfeld equation,
\begin{equation}
  (\p_t{\cal D}-{\cal D}^2)\phi=\ui k\Rey (\p_z^3\psi_0-\p_z\psi_0{\cal D})
  \phi,\quad {\cal D}=\p_z^2-k^2
\label{rvwb8}
\end{equation}
with non-homogeneous boundary conditions,
\refstepcounter{equation}
$$
  \phi=0, \quad \p_z\phi=-\p_z^2\psi_0 \quad \mbox{at}\quad z=0,\quad
  \eqno{(\theequation{\mathit{a},\mathit{b}})}
  \label{rvwb9}
$$
\refstepcounter{equation}
$$
  \phi=0, \quad \p_z\phi=0 \quad \mbox{at}\quad z=\infty,\quad
  \eqno{(\theequation{\mathit{a},\mathit{b}})}
  \label{rvwb10}
$$

The linear operator in the left hand side of Eq. (\ref{rvwb8})
contains a significant difference with respect to that of Eq. 
(\ref{sch9}), the equation governing the first order stream function
for the case of a slightly curved boundary. Both equations describe
vorticity diffusion, but the biharmonic equation (\ref{rvwb8}) contains
a cut-off through the parameter $k$. Is is precisely this term that
will lead to an asymptotic exponential decay of the velocity field
sufficiently far away from the boundary for any finite $\beta$. The 
exponential decay at long distances arises from the screening
introduced by the simultaneous positive and negative vorticity
produced at the troughs and crests of the wavy wall.

In order to obtain a solution of the Orr-Sommerfeld equation (Eq. 
(\ref{rvwb8})),
we further expand the amplitude $\phi(z,t)$ in power series of 
$k\Rey=2\upi\left(2 \left<u_0^2\right>\right)^{1/2}/L\Om$. This is the
ratio between the amplitude of oscillation of the flow at infinity and 
the wall wavelength. We write,
\begin{equation}
\phi=\phi_0+\ui k \Rey\phi_1+\ldots
\label{rvwb11}
\end{equation}
The function $\phi_0$ obeys the linearized Eq. (\ref{rvwb8}) with  
boundary conditions as in Eqs. (\ref{rvwb9}-\ref{rvwb10}) with
$\phi$ replaced by $\phi_{0}$. The Fourier transform of $\phi_{0}$ is
given by,
\begin{equation}
\hat{\phi}_0(z,\om)=(2)^{-1/2}\hat{u}_0(\om)
\frac{\alpha}{\rho-k}\left(\ue^{-\rho z}-\ue^{-kz}\right),
\label{rvwb_phi01}
\end{equation}
with $\rho\equiv(\alpha^2+k^2)^{1/2}$, and the principal branch of the 
square root is assumed $(\Re\{\rho\}>0)$. Recall that 
$\alpha = ( 1 + \ui{\rm sign}(\om))(\om/2\nu)^{1/2}$. The field
$\phi_0$ describes
vorticity diffusion near the wavy wall caused by the uniform but
oscillatory far field flow. Both the spatial and ensemble averages of 
this flow are zero. However, the flow non-uniformity at this order
induces mean flow at the next order, as it is readily apparent from the 
equation for $\phi_1$,
\begin{equation}
  (\p_t{\cal D}-{\cal D}^2)\phi_1=(\p_z^3\psi_0-\p_z\psi_0{\cal D})\phi_0,
\label{rvwb_phi1}
\end{equation}
with boundary conditions $\phi_1=0, \; \p_z\phi_1=0$ at $z=0,\infty$.
The field $\phi_1$ describes vorticity diffusion forced by the 
nonlinear interaction between $\phi_0$ and $\psi_0$. As was the case 
in Sec. \ref{sec:sch}, we focus on the long time limit of the ensemble
average of Eq. (\ref{rvwb_phi1}) $\phi_1^{(s)} = \lim_{t \rightarrow 
\infty} \left<\phi_1\right> = \chi + \ucc$, where $\chi$ is given by, 
\begin{equation}
  {\cal D}^2\chi=-\frac{1}{2}G(z,\beta),
\label{rvwb12}
\end{equation}
with,
\begin{eqnarray}
  G(z,\beta)=\int_{0}^{\infty}\ud\om\,P(\om,\beta)\,Q(z,\om,\beta),\nonumber\\
  Q(z,\om,\beta)=\alpha(\rho+k)(2\ue^{-(\alpha^*+\rho)z}- 
  \ue^{-(\alpha^*+k)z}-\ue^{-\rho z}).
\label{rvwb13}
\end{eqnarray}
The corresponding boundary conditions are homogeneous,
$ \chi=0, \; \p_z\chi=0$ at $z=0,\infty$. The solution is,
\begin{eqnarray}
\chi(z,\beta) & = &\int_{0}^{\infty}\ud\om\,P(\om,\beta)\hat{\chi}(z,\om),
\nonumber\\
\hat{\chi}(z,\om) & = &
A_1\ue^{-\rho z}+A_2\ue^{-(\alpha^{*}+\rho)z}+A_3\ue^{-(\alpha^*+k)z}+
(B_1+z B_2)\ue^{-kz}
\label{rvwb_chi}
\end {eqnarray}
where the functions $A_i,\,B_i$ depend on frequency and wavenumber,
\begin{eqnarray}
A_1=D/\alpha^4,\quad A_2=D/(2\alpha^2\rho^2),\quad 
A_3=-D/(\alpha^2(\alpha^*+2k)^2,\quad D=\alpha(\rho+k)/2,\qquad\nonumber\\
B_1=-(A_1+A_2+A_3),\quad B_2=(\rho-k)A_1+(\rho+\alpha^*-k)A_2 + 
\alpha^*A_3.\qquad
\label{rvwb_chi_const}
\end{eqnarray}
Therefore the stationary part of the averaged first order stream
function is given by,
\begin{equation}
\psi_1^{(s)}=\ui k \Rey\hat{\eta}\uexp(\ui kx)(\chi+\chi^*)+\ucc
\label{rvwb17}
\end{equation}
This solution shows that $\psi_1^{(s)}$
has a phase advance of $\upi/2$ with respect to the wall profile, and hence 
the flow in the vicinity of the boundary is directed from trough to crest 
($\p_z^2(\chi+\chi^*)|_{z=0} > 0 )$.
By shifting the coordinate system along the $x$ axis we can change the
phase of the wall profile so as to make it a simple cosine function.
We consider $\eta(x)=\eps\ucos(kx)$ in what follows.

Following \cite{re:lyne71}, we now proceed to study the limits of
$k$ large and small, while $k\Rey \ll 1$. For $k \gg 1$ the wavelength of the 
boundary profile is much smaller that 
the thickness of the viscous layer. In this case, a boundary layer 
appears of characteristic thickness $1/k$. Screening between regions
producing positive and negative vorticity occurs over a distance
much smaller that the Stokes thickness based on the frequency of
oscillation. The net vorticity does not diffuse even to
distances of order $z\sim\cal {O}$(1), hence
giving rise to exponential decay with an $\uexp(-kz)$
factor. The region in which the stream function is not exponentially
small depends on $Z = kz$. The explicit form of $\psi_1^{(s)}$ may be 
obtained by direct expansion of the solution (\ref{rvwb17}) in power 
series of $1/k$, keeping $Z$ fixed. 
The leading contribution to the steady part of 
the tangential component of the velocity is given by,
\begin{equation}
u_1^{(s)}=k\p_Z\psi_1^{(s)}\sim-\frac{\Rey}{24k^2}\usin(kx)\,\ue^{-Z}Z(6-Z^2)
\label{wavy_klarge}
\end{equation}
The  boundary layer comprises two recirculating cells per wall 
period, located  within $0<z\lesssim1/k$. The separation point is given 
by $z^2=6/k^2$.

\begin{figure}
  \vspace{1.5pc}
  \centerline{\epsffile{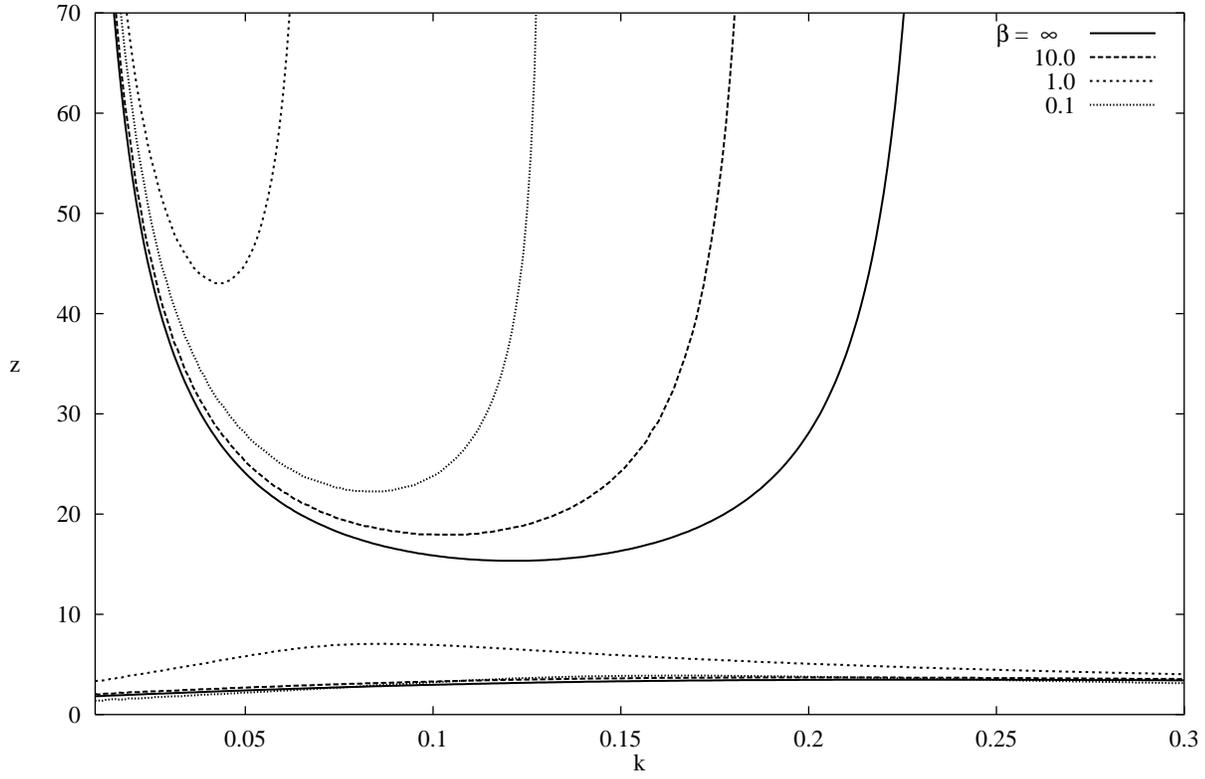}}
\caption{Separation points of the boundary layer over the wavy boundary
as a function of its wavenumber $k$, for different values of $\beta$.
The separation points are the loci of zero tangential velocity. At fixed
$k$, the location of the second separation point depends strongly on $\beta$.}
  \label{fig:wavy_uzero}
\end{figure}

In the opposite limit of $k \ll 1$ one formally recovers the Schlichting 
problem in that the characteristic longitudinal length scale is much 
larger than the Stokes layer thickness. There is one fundamental
difference, however, which can be seen from the solution,
Eq. (\ref{rvwb_chi}). It has two contributions: the first one is
proportional $A_i$, arises from the particular solution,and serves to
balance the non-homogeneity in Eq. (\ref{rvwb12}). This contribution
decays within the Stokes layer. The second one is proportional to
$B_i$, and arises from the general solution of the homogeneous part of 
the equation. This contribution decays over the stretched scale
$Z$. It turns out that this second
contribution introduces an additional separation of the composite
boundary layer when $0<k\leq0.23$. The lines of zero value of the
longitudinal component of the steady velocity profile
are shown Fig. \ref{fig:wavy_uzero}. The location of the second
separation point is entirely determined by that part of the solution
that is proportional to $B_{i}$, and occurs at $z \sim {\cal{O}} (1/k) \gg 1$.

The steady velocity may be obtained by expanding the 
exact solution (Eq. (\ref{rvwb_chi})) in power series of $k$, first
keeping $z$ fixed (inner solution, $u_{1i}^{(s)}$),
and second keeping $Z$ fixed (outer solution $u_{1o}^{(s)}$).
To leading order, we find,
\begin{eqnarray}
u_{1i}^{(s)}(z') & \sim &  -\Rey k^2\usin(kx)
  \left\{ \big[\frac{z'}{2}(\usin(z')-\ucos(z'))+2\usin(z')+\frac{1}{2}
\ucos(z')\big]\ue^{-z'}
  \right. \nonumber\\
&& \left.\mbox{}\hspace{70pt}+  
   \frac{\ue^{-2z'}}{4} -\frac{3}{4}\astrut\right\},\quad 
z' = z/2^{1/2}\sim {\cal{O}}(1),\\
u_{1o}^{(s)}(Z) & \sim & -\Rey k^2\usin(kx)\big[\frac{3}{4}(Z-1)\ue^{-Z}\big],
\qquad Z\sim {\cal{O}}(1) .
\label{wavy_io_sols}
\end{eqnarray}
The solution for the inner and outer steady velocities was already obtained by
\cite{re:lyne71} by a conformal transformation technique. We further note that
the inner and outer solutions can now be matched by requiring that
$u_{1i}^{(s)}(\infty)=u_{1o}^{(s)}(0)=3/4\Rey k^2\usin(kx)$. Hence it
is possible to construct a uniformly valid solution by adding the
inner and outer solutions, and subtracting the first term of the
inner expansion of the outer solution. We find, 
\begin{eqnarray}
u_{1c}^{(s)}(z') & \sim &  -\Rey k^2\usin(kx)
  \left\{ \big[\frac{z'}{2}(\usin(z')-\ucos(z'))+2 \usin(z') +
  \frac{1}{2}
  \ucos(z')\big]\ue^{-z'} \right. \nonumber\\
&& \left.\mbox{} \hspace{70pt}+  
   \frac{\ue^{-2z'}}{4} + \frac{3}{4}(2^{1/2}kz'-1)\ue^{-2^{1/2}kz'}
  \astrut \right\}
\label{wavy_csol}
\end{eqnarray}

\begin{figure}
  \vspace{1.5pc}
  \centerline{\epsffile{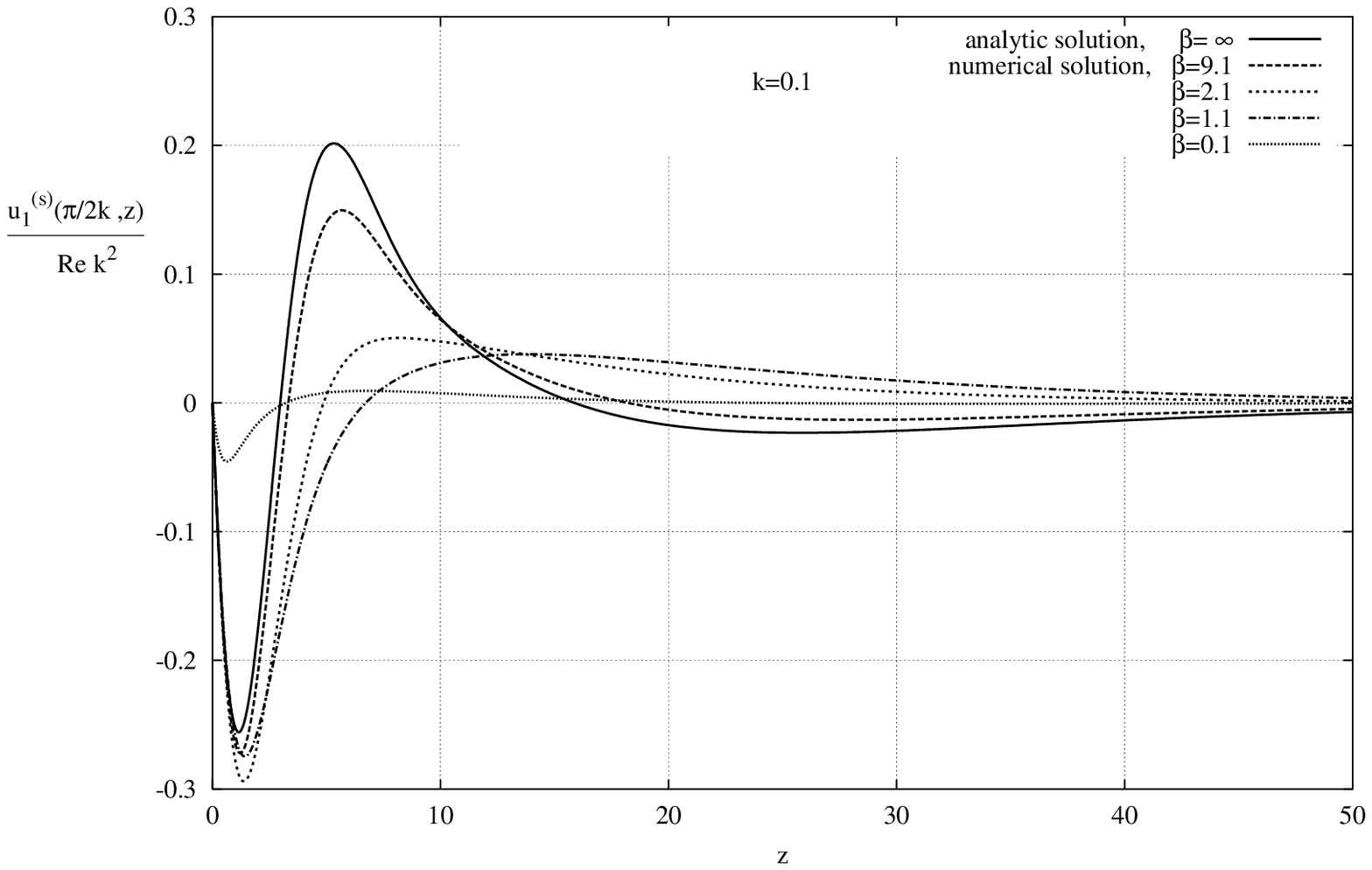}}
\caption{Tangential component of the mean stationary velocity as a function
of $z$ for $k=0.1$ and a range of values of $\beta$. The case $\beta=\infty$
corresponds to analytic solution obtained by Lyne. The other
curves are the numerical solutions of the boundary value problem
defined by Eq. (\ref{rvwb12}) and corresponding boundary conditions.}
\label{fig:wavy_ks}
\end{figure}

\begin{figure}
  \vspace{1.5pc}
  \centerline{\epsffile{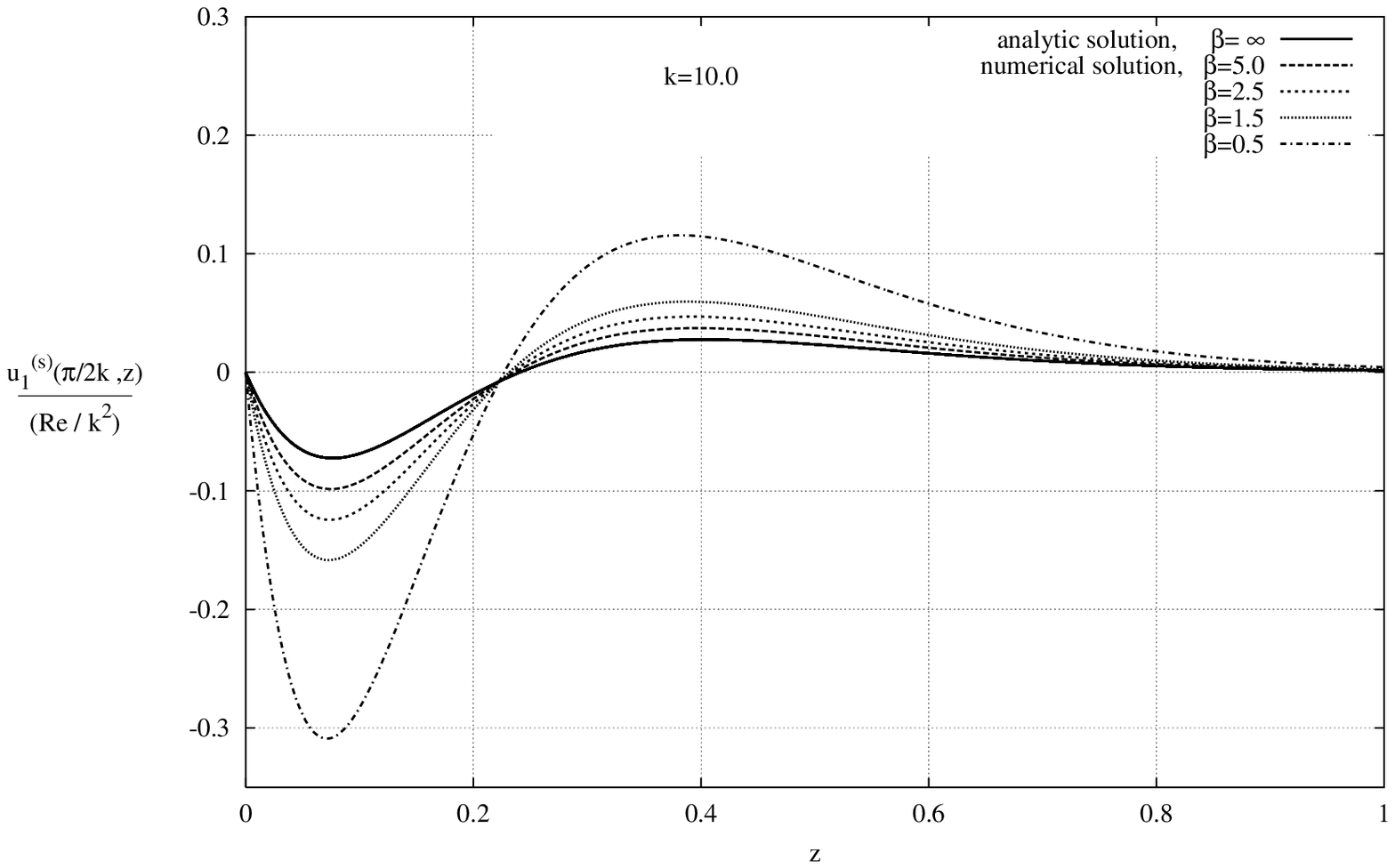}}
\caption{Tangential component of mean stationary velocity as a function
of $z$ for $k=10$ and a range of values of $\beta$. The case $\beta=\infty$
corresponds to analytic solution obtained by Lyne. The other
curves are the numerical solutions of the boundary value problem
defined by Eq. (\ref{rvwb12}) and corresponding boundary conditions.}
\label{fig:wavy_kl}
\end{figure}

We now turn to a numerical study of the case of finite $\beta$.
The boundary value problem (\ref{rvwb12}) has been solved numerically
by using a multiple shooting method for non stiff and linear boundary 
value problems (\cite{re:mattheij84}). The method has the advantage
that the necessary intermediate shooting points are determined by the method itself,
and that it can give the solution on a preset and nonuniform grid of points.
The code was tested on the analytically known solution of the deterministic
limit, Eqs. (\ref{rvwb_chi}) and (\ref{rvwb_chi_const}). Our results are
summarized in Figs. \ref{fig:wavy_uzero}, \ref{fig:wavy_ks} and \ref{fig:wavy_kl}.

Figure \ref{fig:wavy_uzero} shows the separation points of the stationary
velocity as a function of the boundary wavenumber $k$ for a range of values of
$\beta$, including for reference the deterministic limit of $\beta = \infty$
(separation points are defined to be the zeros of the stationary tangential velocity).
This figure shows that for fixed $k$, the mean stationary velocity field may 
be comprised
of two or four recirculating cells per wall period depending on $\beta$. 
The first separation point is largely independent of $\beta$, whereas
the deviation of the second relative to its value in the monochromatic
limit is proportional to $\beta/\upi(1+\beta^2)$, the value of $P(0,\beta)$.  

Our results in the limit $k\ll1$ are presented in Fig. \ref{fig:wavy_ks}, where
profiles of tangential component of the mean stationary velocity 
are plotted for $k=0.1$ and different values of $\beta$. At large
$\beta$ (close to the monochromatic limit) the flow is comprised of four
recirculating cells per boundary period. Upon decresing $\beta$, the
second separation point moves to infinity (see also Fig.
\ref{fig:wavy_uzero}), so that beyond some critical value of $\beta$,
only two recirculating cells remain. Further decrease in $\beta$ results
in the reappearance of the second separation point at infinity, which
then continues to move towards decreasing $z$. The intensity of the
recirculating modes does not change monotonically with $\beta$ as we
further discuss below. In the opposite limit of $k \gg 1$ 
(Fig. \ref{fig:wavy_kl} shows the case $k=10.0$; note that
$u_1^{(s)}$ is now normalized by $\Rey/k^2$), the qualitative structure of 
the flow is largely independent of $\beta$. The streaming flow has
two recirculating cells per wall wavelength, and their intensity increases
monotonically with decreasing $\beta$.

The complex dependence of the flow on $\beta$ and $k$ can be
qualitatively understood from the interplay between the width 
of the power spectrum (given by $1/\beta$), the viscous damping of each 
elementary excitation that depends on its frequency, and the penetration 
depth of the flow field which is primarily dictated by the boundary wavelength. 
For small $k$, large frequency modes are damped close to the boundary and
do not penetrate much into the recirculating layers. Reducing $\beta$
introduces high frequency components into the driving terms at first
order, but they are dynamically damped.
At the same time, the power in the dominant frequency components (around
$\Omega$) decreases. Overall, a decrease in $\beta$ then leads to a decrease in
recirculation strength. As $k$ increases, larger
frequencies contribute to the flow over the entire range of the
recirculating cells. Decreasing $\beta$ decreases the strength of the
dominant components, but increases the range of high frequencies that
can contribute to the flow. From Eq. (\ref{rvwb12}) one can show that the
driving contribution from higher frequencies which is contained in $Q$ 
increases faster
with frequency than the decreasing weight given to them by the power
spectrum $P(\omega,\beta)$. Consequently, decreasing $\beta$ (which
amounts to moving towards the white noise limit) leads to increasing
amplitude of the recirculation.

In summary, for any value of $\beta$, finite or infinite, the vorticity 
produced by vibration of the wavy boundary does not penetrate into the bulk
farther than a distance of order of the wavelength of the boundary. 
However, there are qualitative differences with the deterministic
limit in the character of the flow within that layer. In particular,
the structure and the intensity of the stationary secondary flow
strongly depend on $\beta$.

\section{Summary}

We have addressed the flow induced by a randomly vibrating solid
boundary in an otherwise quiescent fluid. This analysis has been 
motivated by the random residual acceleration field in which microgravity
experiments are conducted. The salient features of the flow are
summarized below.

When the solid boundary is planar, the flow field averages to zero
(the average velocity of the boundary has been taken to be zero in all
cases investigated), but its variance decays algebraically with
distance away from the wall. This dependence follows from a
non vanishing power spectrum of the boundary velocity at zero
frequency. Introducing a low frequency cut-off in the power spectrum
leads back to the classical exponential decay, with a rate that is
determined by the cut-off frequency, Eq. (\ref{plane4_cut}). 
The amplitude of the decaying variance depends explicitly on the 
correlation time of the boundary velocity, $\beta = \Omega \tau$,
where $\Omega$ is the dominant angular frequency of the power spectrum
of the boundary velocity, and $\tau$ is inverse spectral width ($\tau$
is the correlation time of the boundary velocity).

If the solid boundary is curved, steady streaming is generated in
analogy with the classical analysis of Schlichting. The stationary part
of the ensemble average of the secondary velocity is nonzero, even though
the boundary velocity averages to zero. In this case, we find that the 
leading contribution to the average stationary velocity diverges 
logarithmically with distance away from the boundary. 
In analogy to the planar case, the introduction of a low frequency cut-off
in the power spectrum of the boundary velocity changes the asymptotic behavior
qualitatively. The average stationary velocity asymptotes now to a constant, 
given by Eq. (\ref{sch_cut_asym2}). The asymptotic velocity explicitly 
depends on $\beta$ and logarithmically on the cut-off frequency. This asymptotic
behavior is not reached until a length scale of the order of the Stokes layer 
thickness that is based on the cut-off frequency.

We have finally analyzed the case of a periodically modulated solid
boundary in the limit in which the scale of the wall modulation is
small compared to the thickness of the Stokes layer, and also when the
spatial amplitude of the boundary oscillation is small compared with
the wavelength of the wall profile. Cancellation of vorticity
production over the wall boundary leads to exponential decay of the
fluid velocity away from the boundary, with a decay length which is
proportional to the wall wavelength, even if the zero
frequency value of the power spectrum of the boundary velocity is
nonzero. If the boundary wavelength is much larger than the Stokes
layer thickness, we find steady streaming in secondary flow with two or
four recirculating cells per wall period depending on $\beta$.
On the other hand, if the wavelength is
much smaller than the Stokes layer thickness, only two recirculating
cells are formed regardless of the value of $\beta$. Somewhat unexpectedly,
the intensity of the recirculation can both increase or decrease with $\beta$.

\begin{acknowledgments}
This research has been supported by the Microgravity Science and Applications
Division of the NASA under contract No. NAG3-1885, and also in part
by the Supercomputer Computations Research Institute, which is
partially funded by the U.S. Department of Energy, contract No.
DE-FC05-85ER25000.
\end{acknowledgments}

\newpage
\bibliographystyle{$HOME/tex/jfm}
\bibliography{dv1}

\newpage
\newpage
\newpage
\newpage
\begin{figure}
  \vspace{1.5pc}
  \centerline{\epsffile{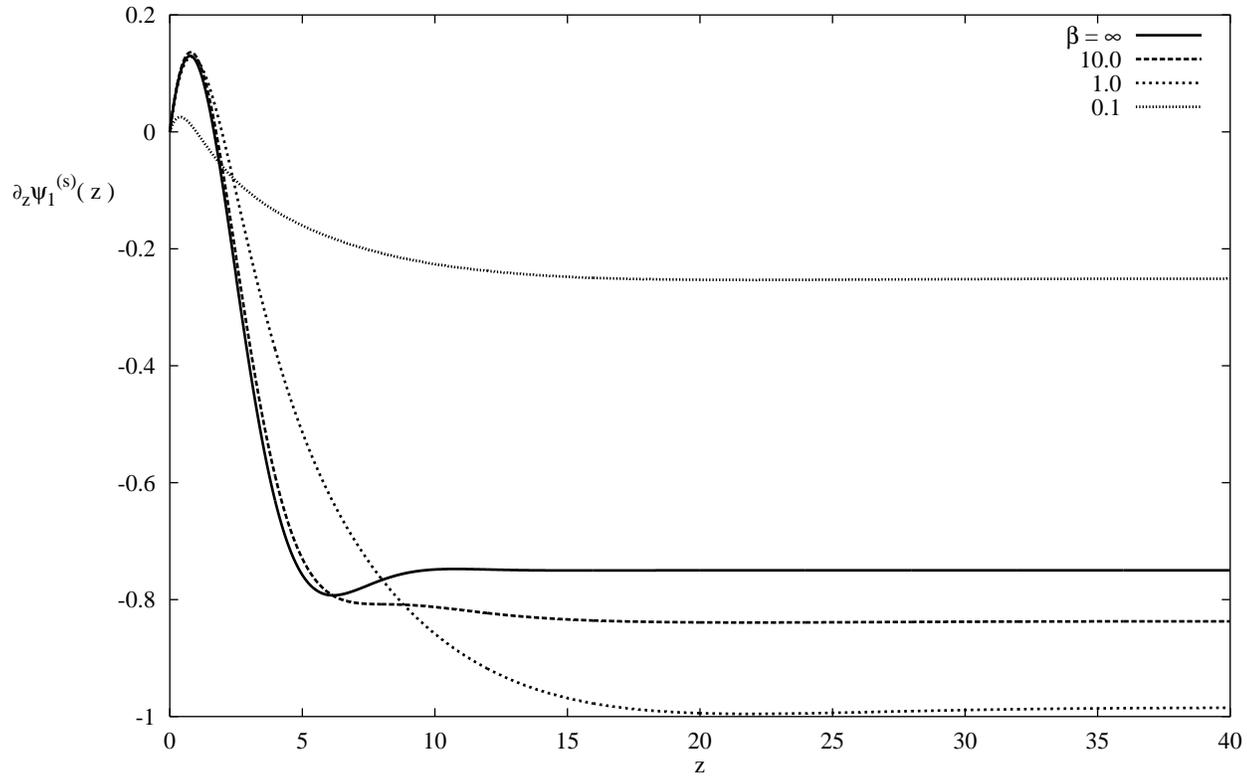}}
\caption{Stationary first order velocity as a function of distance for a
range of values of $\beta$. The power spectrum of the boundary velocity
has a low frequency cut-off at $\om_c=0.05$. The velocity asymptotes to
a constant that depends on the value of $\beta$.}
  \label{fig:usc_cut_prof}
\end{figure}
\newpage
\newpage
\newpage
\newpage
\newpage

\end{document}